\documentclass[aps,prd,twocolumn,showpacs,preprintnumbers,amsmath,amssymb,a4paper]{revtex4-1}
\usepackage{graphicx}
\usepackage{verbatim}
\usepackage{float}
\usepackage{epsf}
\usepackage[usenames,dvipsnames]{xcolor}
\newcommand{\beq}{\begin{equation}}
\newcommand{\eeq}{\end{equation}}
\newcommand{\bea}{\begin{eqnarray}}
\newcommand{\eea}{\end{eqnarray}} 
\newcommand{\nn}{\nonumber}
\begin{document}
%\preprint{Draft.13}
\title{Natural Emergence of Cosmological Constant and Dark Radiation from Stephenson-Kilmister-Yang-Camenzind Theory of Gravity}
\author{Pisin Chen$^{1,2,3,4}$}
\email{pisinchen@phys.ntu.edu.tw}\emph{}
\author{Keisuke Izumi$^{3}$}
\email{izumi@phys.ntu.edu.tw}\emph{}
\author{Nien-En Tung$^{1,3}$}
\email{d96222003@ntu.edu.tw}
\affiliation{%
1. Department of Physics, National Taiwan University, Taipei, Taiwan 10617\\
2. Graduate Institute of Astrophysics, National Taiwan University, Taipei, Taiwan 10617\\
3. Leung Center for Cosmology and Particle Astrophysics, National Taiwan University, Taipei, Taiwan 10617\\
4. Kavli Institute for Particle Astrophysics and Cosmology, SLAC National Accelerator Laboratory, Stanford University, Stanford, CA 94305, U.S.A.
}%

\date{\today}

\begin{abstract}
We show that the Stephenson-Kilmister-Yang (SKY) equation combined with Camenzind's matter current term naturally provides the cosmological constant and dark radiation as integration constants of the SKYC field equation.
To characterize the property of the dark radiation, we develop a method to separate it from the ordinary radiation.
 We found a special property of Camenzind's matter current, namely that the solution space for radiation in fact belongs to that of the vacuum solution of SKY equation. We also found that his matter current does not obey the conservation condition suggested by Kilmister. 
Finally, we  discuss the possible role of dark radiation emergent from the SKYC theory in recent cosmic-microwave-background observations and its implications to the inflation scenario.
\end{abstract}

\maketitle

\section{Introduction}
Various attempts have been made throughout the last century to unify all fundamental interactions. One pioneering effort was made by Weyl in 1918 \cite{Weyl:1918ib,Weyl1919}, where he assumed that all physical laws should be invariant under conformal transformation. This seminal viewpoint introduced by Weyl is now known as the gauge invariance. Based on the principle of gauge invariance, Weyl reformulated the connection in Riemannian geometry  to unify gravity and electromagnetism. 

The gauge theory of Yang and Mills has inspired a new type of relationship between geometry and physics \cite{PhysRev.96.191}. Such a theory can be considered as a vector bundle on a Riemannian manifold \cite{Cho1975,Trautman197029,PhysRevD.13.235}, where a section of a vector bundle corresponds to a matter field and a connection on the Riemannian manifold corresponds to a gauge field. A gauge theory can therefore be recognized as a functional (called Yang-Mills functional) that acts on a metric connection on a vector bundle. Yang-Mills functional is also invariant under gauge transformation. Thus, Yang-Mills theory exhibits a close correspondence with the vector bundle theory in differential geometry. 

We now know that all fundamental interactions: the electromagnetic, the weak, and the strong, can be described in the language of gauge theory except general relativity (GR). The pioneer works to formulate Einstein's GR into a gauge theoretical framework started with Utiyama, who suggested that GR can be written in the language of the gauge theory if the symmetry group is chosen either as the Poincar{\'e} group or the translational gauge group \cite{Utiyama1956,kibble:212,PhysRevLett.36.59,PhysRevD.14.2521}. 

 Yang proposed his gauge theory of gravity based on the $GL(n)$ group and identified the affine connection as the gauge potential \cite{PhysRevLett.33.445}, with the action for the pure space proportional to $R^{\mu\nu\sigma\lambda}R_{\mu\nu\sigma\lambda}$, where $R_{\mu\nu\sigma\lambda}$ is the Riemann tensor. Prior to Yang's proposal, Stephenson, Kilmister and Newman had derived a similar equation but with additional constraints from a second-ordered curvature Lagrange density \cite{Steph1958,Kilmister1961}. We shall call their field equation the SKY equation. The SKY equation reproduces all solutions of the vacuum Einstein equation~\cite{Steph1958,Kilmister1961,PhysRevLett.34.507,PhysRevLett.35.320,PhysRevLett.34.1114,Barrett1977278}. Later Camenzind proposed a matter current term for the SKY field equation \cite{Camenzind1975}, although it was not deduced from an action. This task was fulfilled by Cook in 2009 \cite{Cook:2009mx}. We shall call the complete theory that includes both pure space and matter contributions the SKYC gravity. A renormalizable quantum theory of gravity based on the Einstein-Hilbert action is known to be difficult to attain \cite{citeulike:6102261,PhysRevD.10.401,0264-9381-9-4-006}.
Being a gauge theory, a quantum gravity based on the SKYC theory might stand a better chance to be renormalizable \cite{PhysRevLett.33.445,PhysRevD.16.953,springerlink:10.1007/BF00760427}. On the other hand, the high-order, quadratic contraction of Riemann tensor without the protection of the Gauss-Bonnet condition will necessarily render the ghost problem. One may, however, work around the ghost problem by expelling the negative pole to the UV limit, say to the Planck scale, and regarding the SKYC gravity as an effective theory \cite{Chen:2010at}. 
  
 There has been the longstanding cosmological constant problem \cite{RevModPhys.61.1}, The nature of the cosmological constant (CC) term introduced by Einstein to his field equation is a priori undefined. On the other hand the quantum vacuum energy satisfies  the properties of the CC, yet its value is about 124 orders of magnitude larger than the critical density of the universe, which is comparable to what is required for CC to explain the observed accelerating expansion of the universe \cite{Perlmutter:1998np,Riess:1998cb}. From the mathematical consideration, the CC term cannot be removed from the Einstein-Hilbert action. The SKYC equation is second order in terms of the affine connection. While the metric is a priori not a dynamical variable, in order to reduce the SKYC equation to the Einstein equation one define the relation between the connection and the metric in the usual way. As a result the SKYC equation can be redressed as a 3rd order differential equation of the metric where the CC term is necessarily absent. The CC term, however, is recovered as an integration constant in reducing the SKYC equation to Einstein equation. In this approach CC is no longer arbitrary but determined by the boundary condition of the universe, which is geometrical in nature and has nothing to do with the quantum vacuum energy. For example it has been proposed that the underlying geometry is the de Sitter space and this integration constant has been associated with the radius of curvature of the de Sitter space \cite{Cook:2009mx,SQJian2009,Chen:2010at}. Therefore the SKYC formulation of gravity may provide a solution to the CC problem. It goes without saying, of course, that a big part of the problem still remains: why does the quantum vacuum energy {\it not} contribute to gravitation in the first place? This issue is beyond the scope of this paper and we will not dwell on it further.
  
 Aside from the CC problem, recent observation data indicate that there is an apparent increase of the effect number of neutrino flavors, $N_{\rm eff}$, between the big bang nucleothynthesis (BBN) and the cosmic microwave background (CMB) epochs \cite{wmap2010,ACT2010,SPT2011,arXiv:1109.2767,planck}. The standard model predicts $N_{\rm  eff}=3.04$ at the epoch where the universe was dominated by photons and neutrinos after the electron-positron annihilation. However, recent experimental data suggest a larger $N_{\rm eff}$ \cite{wmap2010,ACT2010,SPT2011,arXiv:1109.2767,planck}. It has been suggested that the difference, $\Delta N_{\rm eff}\equiv N_{\rm eff}-3 \simeq 1$ at nearly $2\sigma$ level, may indicate the existence of an additional, heretofore unobserved ``dark radiation (DR)" density. On theoretical side, this additional radiation was also called upon by the brane-world inspired and other cosmological models \cite{Mukohyama:1999qx,Ryusuke20120814,Nakayama:2010vs,Fischler:2010xz,Kawasaki:2011ym,Hasenkamp:2011em,Blennow:2012de}. It happens that another integration constant in the SKYC theory has exactly the required character of dark radiation. We note, however, as has been pointed out recently by Birrell et al. \cite{Birrell2013}, that the fact that neutrinos have rest mass and that their distributions are non-thermal under free-steaming, can well explain such an increase of $N_{\rm eff}$ without the need to invoke dark radiation. With that in mind, the dark radiation arisen from SKYC gravity is nonetheless a free bonus at our disposal subject to observational constraints.
  
The purpose of this paper is to examine explicitly the salient features, in particular the CC and dark radiation, of the SKYC gravity mentioned above. To do so, we derive the Friedmann-Lema\^{i}tre-Robertson-Walker (FLRW) equation associated with the SKYC gravity with integration constants. We identify one integration constant as CC and the other as dark radiation. Subsequently, we develop a method to investigate the property of dark radiation and demonstrate that the density of dark radiation is a constant on the constant-time hypersurface in every metric.

 This paper is arranged as the following. In Sec.~\ref{sec:skyc}, we review the SKYC gravity. In the next section, the SKYC field equation on the FLRW metric is derived and investigated.  We discuss the property of nullity in Camenzind's matter field in the perfect fluid model. In Sec.~\ref{sec:inhomogeneous}, we characterize the properties of the dark radiation. In Sec.~\ref{sec:snd}, we summarize and discuss our findings. In the appendix, we  discuss the problem with Cook's theory, which intends to give rise to the SKYC equation from the action level.

\section{SKYC Equation}
\label{sec:skyc}

In a non-abelian gauge theory, the field strength is defined as
\beq
F_{\mu\nu}=\frac{i}{g}[D_\mu ,D_\nu],
\label{feq}
\eeq
where $D_\mu=\partial_\mu-igA_\mu$ and $A_\mu$ is the gauge potential.
The field strength $F_{\mu\nu}$ can therefore be expressed in terms of the gauge potential $A_\nu$ as
\beq
F_{\mu\nu}=\partial_\mu A_\nu - \partial_\nu A_\mu - ig[A_\mu , A_\nu].
\eeq
When connecting Riemmanian geometry with the non-abelian gauge theory, 
the Christoffel symbol, $\Gamma^{\lambda}_{\mu\nu}$, and the Riemann tensor, $R_{\mu\nu\sigma\lambda}$, play the role of the gauge potential $A_\mu$ and the field strength $F_{\mu\nu}$, respectively. Motivated by this gauge connection, Stephenson, Kilmister and Newman investigated an alternative theory of gravity and obtained the following equation of motion, 
\begin{equation}
\nabla_{\nu}{R^{\nu}_{~\mu\alpha \beta}}=0,
\label{GaugeFieldEq1}
\end{equation}
which is  a higher-order derivative with respect to the metric \cite{Steph1958,Kilmister1961}.
Invoking the second Bianchi identity, it can be readily verified that Eq.(\ref{GaugeFieldEq1}) is equivalent to
\beq
\nabla_{\mu}R_{\alpha \beta}-\nabla_{\beta}R_{\alpha \mu}=0.
\label{SKNYeq}
\eeq
In fact, Eq.(\ref{GaugeFieldEq1}) is exactly Yang's gravitational field equation for pure space under the $GL(n)$ gauge symmetry \cite{PhysRev.96.191}.

To complete the gravitational field equation, 
 Kilmister introduced  the matter current $I$ such that \cite{Kilmister1961}
\beq
\nabla_{\alpha}R_{\beta \mu}-\nabla_{\beta}R_{\alpha \mu}=I_{\alpha \beta \mu} .
\label{Kileq}
\eeq 
He found that $I$ has $20$ independent components under the constraint of the conservation law:
 \beq
 \nabla_{\mu}I^{\alpha  \beta \mu }=0.
 \label{Kicons}
 \eeq 
 However, he did not provide the explicit form for the current density $I$.
 
 Later, Camenzind wrote down a Yang-Mills field equation for $SO(3,1)$ with a current $J$ different from Kilmister's $I$:
  \beq
  \nabla_{\nu}{R^{\nu}_{~\mu\alpha \beta}}= 
  8\pi G J_{\alpha 
\beta\mu},
\label{CamEq}
  \eeq
where the current density $J^{\alpha 
\beta\mu}$ has the form:
  \bea
  \label{defJ}
  J^{\alpha 
\beta\mu}=&\nabla^{\beta}& \left( {T}^{\mu\alpha}-1/2g^{\mu \alpha}T^{\lambda}_{~\lambda} \right)  \\ 
&-&\nabla^{\alpha} \left({T}^{\mu\beta}
-1/2g^{ \mu\beta}T^{\lambda}_{~\lambda} \right). \nonumber
  \eea
Here $T^{\mu\alpha}$ is the energy-momentum tensor \cite{Camenzind1975}. 
Nevertheless, an action associated with this matter current remained lacking.
Aspired by the analogy with Maxwell's theory, Cook later proposed an action term for the matter current, $\Gamma_{\alpha\beta}^{\mu}J^{\alpha\beta}_{\mu}$, from which the SKYC field equation can be derived \cite{Cook:2009mx}.
It can be verified that solutions of
Eq.(\ref{CamEq}) cover the entire solution-space of the Einstein equation with source \cite{PhysRevD.16.2438}.

\section{SKYC Equation in FLRW Metric}
\label{sec:flrw}
 In this section we will use the FLRW metric to derive a modified Friedmann equation from the SKYC field equation, Eq.(\ref{CamEq}), and discuss the nullity of Camezind's current density in the radiation-dominated case. 

\subsection{EOM in  Homogeneous Universe}
\label{sec:flrw:1}
 We consider a homogeneous   and isotropic universe and use  the FLRW metric to study Eq.(\ref{CamEq}). The  metric is:
\beq
ds^2=-dt^2+a^2(t)\left(\frac{dr^2}{1-kr^2}+r^2 d\Omega^2 \right),
\eeq
where $a(t)$ is the scale factor.
From the symmetry property of the FLRW metric, the energy-momentum tensor $T_{\mu\nu}$ can be written as
\beq
T_{\mu\nu}=(\rho+p)u_\mu u_\nu + p g_{\mu\nu},
\label{tensor}
\eeq
where  $u^\mu = (\partial/\partial t) ^{\mu}$.
In Eq.(\ref{CamEq}), we assume  $T_{\mu\nu}$ to satisfy the conservation law:
\beq
\nabla_\mu T^{\mu\nu}=0,
\label{conserv}
\eeq
which in turn leads to the usual equation
\beq
\dot{\rho}=-3H(\rho+p),
\label{conserv1}
\eeq
where $H\equiv\dot{a}/{a}$.

The only nontrivial components of  Eq.(\ref{CamEq}) are:
\beq
\nabla_\nu R^{\nu}_{~i0i}=8\pi J_{0ii}, \\
\eeq
where no summation over the repeated index $``i"$ occurs.
Other components  vanish due to the symmetry of the metric.
The expression for $T_{\mu\nu}$, Eq.(\ref{tensor}), leads to 
\beq
a^2\dddot{a}+a\dot{a}\ddot{a}-2\dot{a}^3-2k\dot{a}=8\pi G\left( \frac{1}{2}a\left(\dot{\rho}-\dot{p}\right)+\dot{a}(\rho+p) \right),
\label{radsol22}
\eeq
which can also be expressed as
\beq
\ddot{H}+4H\dot{H}-2k\frac{\dot{a}}{a^3}=8\pi G\left(\frac{1}{2}\left(\dot{\rho}-\dot{p}\right)+H(\rho+p)\right). \\
\label{Fried7}
\eeq
Eq.(\ref{Fried7}) is the modified Friedmann equation in SKYC theory.
By integrating Eq.(\ref{Fried7}) with Eq.(\ref{conserv1}), we can arrive at
\beq
\dot{H}+2H^2+\frac{k}{a^2}=\frac{8\pi G}{2}\left(\frac{1}{3}\rho-p \right)+2C_1 ,
\label{Heq1}
\eeq
where $C_1$ is an integration constant.
By performing another integration to Eq.(\ref{Heq1}), one obtains
\beq
H^2=\frac{8\pi G}{3}\rho -\frac{k}{a^2}+ C_1+\frac{C_2}{a^4}
\label{Heq2}.
\eeq
This is the Friedmann equation for the SKYC theory.

Eq.(\ref{Heq2}) can be interpreted as follows. In {\color {blue}a} FLRW universe,  $C_1$ plays the role of the CC term.   
As will be shown in the following section,  the constant $C_1$ plays the same role as the CC even in an inhomogeneous universe; this fact  has been shown by Cook \cite{Cook:2009mx}. (Our $C_1$ is identical to the trace part of $X_{\mu\nu}$ in his article.) 
The term involving $C_2$, on the other hand, scales the same as radiation. Since it does not couple with any matter except gravity, it can be identified as the dark radiation. We note that although the $C_2$ term is not induced by the matter current, it  does affect the evolution of the universe. %\cite{a}.

\subsection{Nullity in Current Density Tensor}
\label{sec:flrw:2}
In this subsection we show that Camenzind's current density will be null if the equation of state for radiation is invoked. On the other hand, such equation of state belongs to the vacuum  equation in the SKYC gravity. We study this special property here.

The equation of state for radiation is $p=(1/3)\rho$. Based on this, Eq.(\ref{conserv1}) becomes
\beq
\rho_r a^4=E , \nn
\eeq
where $E$ is a constant.
Thus the dynamical equation for $a$ is
\beq
\dot{a}^2=\frac{8\pi GE}{3a^2}-k+ C_1a^2+\frac{C_2}{a^2},
\label{radsol4}
\eeq
where the first term on the RHS represents the contribution from the ordinary radiation. As we have commented in the previous subsection, the $C_2$ term also behaves like radiation. Hence for a perfectly homogeneous universe one cannot distinguish the ordinary radiation from the dark radiation. %energy effective cosmological equation in the brane-world cosmology. 
% It should be non-negative otherwise there is a singularity in 5-dimension \cite{b}. %In WMAP collaboration, total radiation energy  

There is a property about the nullity of the current density tensor to be mentioned.
If one inserts  Eq.(\ref{radsol4}) into the gravitational equation Eq.(\ref{CamEq}),
then the RHS of Eq.(\ref{CamEq}) becomes null, that is,
\beq
\nabla_{\nu}{R^{\nu}_{~\mu\alpha \beta}}=0,
\label{vac_eq}
\eeq
yet the current density is a priori not necessarily zero. To avoid the contradiction, one must impose the ansatz that $J_{\alpha 
\beta\mu}$ equals to zero in the radiation case. Under the assumption of $ p=w \rho,$  we find that both $w=-1$ and $1/3$ will make $J_{\alpha 
\beta\mu}$ vanish. 
 In this situation, Eq.(\ref{radsol4}) now belongs to the set of the vacuum solutions of Eq.(\ref{CamEq}). Such a higher-order gravitational equation  brings  the radiation terms into integration constants of the vacuum equation of  Eq.(\ref{CamEq}) from the current density  $J_{\alpha 
\beta\mu}$. 
This is an interesting property in Camenzind's  current density of the matter.

Although the genesis of the dark components from Eq.(\ref{CamEq}) is interesting, 
it might be a signal of an inconsistency.
This property means that Camenzind's current density for radiation and dark energy cannot be a source of gravity in the exact FLRW universe. 
This result stems from the assumption in Eq.(\ref{defJ}), where the energy density $T_{\mu\nu}$ is introduced, that the energy density $T_{\mu\nu}$ gives rise to the current and that this current satisfies the conservation law. 
A deeper inspection shows, however, that while the Camenzind's current density $J^\alpha_{~\beta\gamma}$ is the current for gravity, yet there is no theoretical origin for the energy conservation law. 
Furthermore, the non-trivial assumption of Eq.(\ref{defJ}) might result in the emergence of the dark components and the fictitious property about nullity because Camenzind's current density is in the form of higher order derivatives, although this assumption is needed to recover GR.

\section{Characteristic of Dark Radiation}
\label{sec:inhomogeneous}
In the previous section, we have seen the fact that one can not separate the effect of the dark radiation from that of the ordinary radiation in the 
  homogeneous universe. In order to distinguish them,  we consider the case of an inhomogeneous universe.

Firstly, we solve the EOM in the inhomogeneous universe.
We define {\color{blue}the} effective energy-momentum tensor as
\begin{eqnarray}
\hat{T}_{\mu\nu} \equiv \frac{1}{8\pi G} G_{\mu\nu} -T_{\mu\nu}.
\label{deftau}
\end{eqnarray}
Because of Eq.(\ref{conserv}), the divergence of this gives
\begin{eqnarray}
\nabla^\mu \hat{T}_{\mu\nu} =0.
\label{contau}
\end{eqnarray}
Substituting (\ref{contau}) into the gravitational EOM, we have
\begin{eqnarray}
\nabla_\alpha \hat{T}_{\beta \gamma} -\nabla_\beta \hat{T}_{\alpha \gamma}
-\frac{1}{2}\left( g_{\beta \gamma} \nabla_\alpha \hat{T} -g_{\alpha \gamma} \nabla_\beta \hat{T} \right)=0,
\label{eqtau}
\end{eqnarray}
where $\hat{T}$ is the trace of $\hat{T}_{\mu\nu}$.
Multiplying $g^{\alpha \gamma}$ by Eq.(\ref{eqtau}), 
we can obtain
\begin{eqnarray}
\nabla_\beta \hat{T}=0,
\end{eqnarray}
where we use Eq.(\ref{contau}). 
Integrating this, we have
\begin{eqnarray}
\hat{T}=4 \Lambda_{\text eff},
\end{eqnarray}
where $\Lambda_{\text eff}$ is a constant. As we have commented at the end of Sec.~\ref{sec:flrw:1}, it plays exactly the role of CC.

We now separate the contributions other than the effective CC from $\hat{T}_{\mu\nu}$:
\begin{eqnarray}
S_{\mu\nu}\equiv  \hat{T}_{\mu\nu}- \Lambda_{\text eff} g_{\mu\nu}.
\end{eqnarray}
We know from Eqs.~(\ref{contau}) and (\ref{eqtau}) that the following equations must be satisfied:
\begin{eqnarray}
&&S=0,\label{traceless}\\
&&\nabla^\mu S_{\mu\nu}=0,\label{cons} \\
&&\nabla_\alpha S_{\beta \gamma} -\nabla_\beta S_{\alpha \gamma}=0.\label{graveq}
\end{eqnarray}
If the form of $S_{\mu\nu}$ is the same as that of radiation fluid on FLRW metric, 
then it can be shown that these equations are satisfied. Therefore, $S_{\mu\nu}$ is related to the $C_2$ term.

In order to see the difference of the effective energy-momentum tensor $S_{\mu\nu}$ from a real radiation fluid, 
we analyze its property on general metric in the form of a fluid:
\begin{eqnarray}
S_{\mu\nu}=(\rho+p)u_\mu u_\nu + p g_{\mu\nu},
\end{eqnarray}
with
\begin{eqnarray}
u_{\mu}u^{\mu}=-1.
\end{eqnarray}
The traceless condition (\ref{traceless}) fixes the relation between $\rho$ and $p$ as
\begin{eqnarray}
\rho=\frac{1}{3 }p.
\end{eqnarray}
Then $S_{\mu\nu}$ becomes
\begin{eqnarray}
S_{\mu\nu} &=& \frac{4}{3}\rho u_\mu u_\nu + \frac{1}{3} \rho g_{\mu\nu} \\
&=& \rho u_\mu u_\nu + \frac{1}{3} \rho h_{\mu\nu},
\end{eqnarray}
where
\begin{eqnarray}
h_{\mu\nu}= g_{\mu\nu} +u_\mu u_\nu
\end{eqnarray}
is the induced metric on the hypersurface which is orthogonal to $u^{\mu}$.

Eq. (\ref{cons}) can be written as
\begin{equation}
\frac{4}{3}u_\beta u^\mu \partial_\mu \rho + \frac{1}{3} \partial_\beta \rho + 
\frac{4}{3} \rho \left(
u_\mu \nabla^\mu u_\beta +u_\beta \nabla^\mu u_\mu 
\right)=0.
\end{equation}
Multiplying $u^\beta$ and ${h_\nu}^\beta$ by the above expression, respectively, we can obtain
\begin{eqnarray}
&&u^\mu \partial_\mu \rho = -\frac{4}{3}\rho\nabla^\mu u_\mu,\label{upart}\\
&&{h_\nu}^\beta\partial_\beta \rho= -4 \rho u_\mu \nabla^\mu u_\nu,\label{hpart}
\end{eqnarray}
where we use 
\begin{eqnarray}
&&u^\beta \nabla^\mu u_\beta = \frac{1}{2} \nabla^\mu \left( u^\beta u_\beta \right) =0,\\
&&{h_\nu}^\beta \nabla^\mu u_\beta =\left( {\delta_\nu}^\beta + u_\nu u^\beta \right) \nabla^\mu u_\beta =\nabla^\mu u_\nu.
\end{eqnarray}
Multiplying $u_\nu$ by Eq.(\ref{upart}) and subtracting it from Eq.(\ref{hpart}), we have
\begin{eqnarray}
\partial_\nu \rho = \frac{4}{3}\rho u_\nu \nabla^\mu u_\mu -4 \rho u_\mu\nabla^\mu u_\nu.
\label{part}
\end{eqnarray}
On the other hand, Eq.(\ref{graveq}) can be written as
\begin{eqnarray}
&&u_\beta u_\gamma \partial_\alpha \rho-u_\alpha u_\gamma \partial_\beta \rho 
+\frac{1}{3} h_{\beta\gamma} \partial_\alpha \rho -\frac{1}{3} h_{\alpha\gamma} \partial_\beta \rho \nonumber\\
&&\qquad
+\frac{4}{3}\rho\bigl(
u_\beta\nabla_\alpha u_\gamma-u_\alpha\nabla_\beta u_\gamma    \nonumber\\
&&\qquad\qquad\qquad\qquad\quad
-u_\gamma \nabla_\alpha u_\beta+u_\gamma \nabla_\beta u_\alpha
\bigr)=0. \label{graveq2}
\end{eqnarray}
Multiplying $u^\gamma u^\beta$ by Eq.(\ref{graveq2}), we have
\begin{eqnarray}
{h_\alpha}^\beta\partial_\beta \rho- \frac{4}{3} \rho u^\beta \nabla_\beta u_\alpha=0.
\end{eqnarray}
Combining it with Eq.(\ref{hpart}), we have
\begin{eqnarray}
&&u_\mu \nabla^\mu u_\nu=0, \label{unablau}\\
&&{h_\alpha}^\beta\partial_\beta \rho =0. \label{hpart0}
\end{eqnarray}
Multiplying ${h_\lambda}^\gamma u^\beta$ by Eq.(\ref{graveq2}), we have
\begin{eqnarray}
-\frac{1}{3}h_{\lambda\alpha}u^\beta \partial_\beta \rho -\frac{4}{3}\rho \nabla_\alpha u_\lambda=0,
\end{eqnarray}
where we have used Eq.(\ref{unablau}).
Combining it with Eq.(\ref{upart}), we find
\begin{eqnarray}
\nabla_\alpha u_\lambda =\frac{1}{3} h_{\alpha \lambda}\nabla^\mu u_\mu.
\label{tracecondition}
\end{eqnarray}
Multiplying $u^\gamma {h_\lambda}^\beta$ and ${h_\epsilon}^\gamma {h_\lambda}^\beta$ by Eq.(\ref{graveq2}), respectively, gives 
\begin{eqnarray}
&&u_\alpha {h_\lambda}^\beta\partial_\beta \rho +\frac{4}{3} \rho \nabla_\alpha u_\lambda
\nonumber\\ 
&&\qquad
-\frac{4}{3} \rho \nabla_\lambda u_\alpha -\frac{4}{3}u_\lambda u^\beta\nabla_\beta u_\alpha=0,
\label{uh}\\
&&\frac{1}{3}h_{\epsilon\lambda}\partial_\alpha \rho -\frac{1}{3} h_{\epsilon\alpha}{h_\lambda}^\beta \partial_\beta \rho-\frac{4}{3} \rho u_\alpha \nabla_\lambda u_\epsilon=0.\label{hh}
\end{eqnarray}
Substituing Eqs.(\ref{part}), (\ref{unablau}), (\ref{hpart0}) and (\ref{tracecondition}) into Eqs.(\ref{uh}) and (\ref{hh}), 
we see that they are automatically satisfied and no additional condition is obtained. 

In summary, we have transcribed the original equations for $S_{\mu\nu}$ into Eqs.(\ref{upart}), (\ref{unablau}), (\ref{hpart0}), (\ref{tracecondition}),  and 
\begin{eqnarray}
\partial_\nu \rho = \frac{4}{3} \rho u_\nu \nabla^\mu u_{\mu}
\end{eqnarray}
that govern the characters of $\rho$ and $u_{\mu}$.
In turn, these conditions constrain the form of the effective energy-momentum tensor. 
Equation (\ref{hpart}) means that energy density must be constant on the hypersurface which is 
orthogonal to $u^{\mu}$.
Therefore, the dark radiation from the effective energy-momentum tensor can affect only the background dynamics and we confirm
the existence of the real radiation fluid from its perturbation.

\section{Conclusion}
\label{sec:snd}

We investigated the SKYC theory of gravity by way of solving its field equation, Eq.(\ref{CamEq}), in a FLRW universe and arrived at a modified SKYC Friedmann equation, Eq.(\ref{Fried7}). Being a higher order derivative equation than that for GR, the SKYC Firedmann equation gives rise to two integration constants when it is reduced to lower order. One of the two is clearly related to the cosmological constant while the other is related to the dark radiation.
We have also demonstrated, in a homogeneous universe, that this dark radiation is indistinguishable from the ordinary radiation. In addition, we pointed out the nullity of the current density $J$ in the radiation case under the FLRW metric.

In order to further pin down the nature of our dark radiation, we turned to a general, inhomogeneous universe and introduced a methodology to look for its possible difference from the ordinary radiation. We solved the EOM in the inhomogeneous universe and constrained the form of the tensor $S_{\mu\nu}$ in Sec.~\ref{sec:inhomogeneous}. 
We found that if $\rho$ is fixed at one point,  then it will be a constant on the hypersurface orthogonal to $u^{\mu}$. 
That means that there does not exist any degree of freedom for the perturbed dark radiation. That is, this SKYC dark radiation only has the zero mode term but has no perturbed term. In contrast, the ordinary radiation can be perturbed and can therefore propagate in all spacetime. We conclude that the SKYC dark radiation is indeed different from the ordinary one. We should like to comment, however, that in our derivation in Section \ref{sec:inhomogeneous}, the expression $S_{\mu\nu}=(\rho+p)u_\mu u_\nu + p g_{\mu\nu}$ is not the most general form. We will pursue a more general expression for it in our future work.

Some comments are in order with regard to the relationship between the SKYC dark radiation and the inflation. The dark radiation $C_2$ term is an integration constant in this theory. That is, it is determined by the initial or boundary condition of the universe. If the SKYC theory is incorporated with inflation, then the density of the dark radiation $\rho_{DR}$ must start from a tiny value. Otherwise the inflation can not be trigged because $\rho_{DR}$ scales as $a^{-4}$ and dominates at early times. At the end of inflation and after $\sim 60$ e-foldings, the scale factor has grown by $\sim 10^{20}$ times. This means that $\rho_{DR}$ must be smaller by 80 orders of magnitude than $\rho_{DR}$ at late times. Being so tiny, we may as well set $C_2$ to zero. On the other hand, if SKYC theory does not include the inflation scenario, then the $C_2$ term can in principle be identified with the dark radiation and be fixed by the observation data. The recent Planck data gives $N_{\rm eff} =3.36^{+0.68}_{-0.64} (95 \%)$ based on the combination of WMAP + highL data  \cite{planck}. However, this fit produces a 2.5 s.d. tension with direct astrophysical measurements of the Hubble constant. Including priors from SN surveys removes this tension and results in $N_{\rm eff}= 3.62^{+0.50}_{-0.48} (95 \%)$.
The larger $N_{\rm eff}$ suggests a need for the dark radiation. Without fusing SKYC gravity with inflation, we fix $C_2$ with the Planck data and find $C_2\sim 5.59\times10^{-32}~ {\rm kg/m^3}$, while including SN priors Planck data gives $C_2\sim 1.02\times10^{-31} ~ {\rm kg/m^3}$.

\section{Acknowledgement}
It is a pleasure to thank R. J. Adler, Je-An Gu, Debaprasad Maity, Yen Chin Ong, Shu-Heng Shao and Yen-Wei Liu  for helpful discussions. Pisin Chen is supported by Taiwan National Science Council under Project No. NSC 97-2112-M-002-026-MY3 and by US Department of Energy under Contract No. DE-AC03-76SF00515. Keisuke Izumi is supported by Taiwan National Science Council (TNSC) under Project No.\ NSC101-2811-M-002-103.

\appendix
\section{Difference between Stephenson's and Yang's Approaches}
Stephenson's gravitational equation is similar to Yang's, but there are some important differences between the two theories. In brief, Stephenson's theory covers less solutions than Yang's even if the connection in Stephenson's theory is identified as the Christoffel symbol. We will discuss this point in more details in this appendix.

In the Palatini formalism, the affine connection and the metric are treated as independent variables and the derived EOMs can determine the relation between the two. For example,  the relation between the connection and the metric can be identified as the condition for the metric compatibility when applying the Palatini formalism to derive EOMs from the Einstein-Hilbert action. Stephenson and Cook obtained EOMs by applying the Palatini formalism to the quadratic curvature Lagrange density for pure gravity without matter.
%Stephenson {\color{red}arrives at similar EOMs to Yang's gravitational equation but not identical.} We will comment this difference between the two theory in the following paragraphs. 
About the matter field in the SKYC theory, Cook introduced his current density tensor at the action level. However, there are problems with his approach,  about which we will comment in Appendix B.

In Yang's theory, the gravitational force is described by the $GL(4)$ gauge field $b_{\mu}^a$. Here the Latin letters stands for the indices of the $GL(4)$ gauge group.
The action is
\beq
S_b=\int dx^4 \sqrt{-g} \left( g^{\mu\alpha}g^{\nu\beta} C_{ac}^{d} C_{bd}^c f_{\mu\nu}^a f_{\alpha\beta}^b \right).
\eeq
where $f_{\mu\nu}^a$ is the field strength:
\beq
f_{\mu\nu}^a=b_{\mu,\nu}^a-b_{\nu,\mu}^a- C_{bc}^a b_\mu^b b_\nu ^c,
\label{bf}
\eeq
and $C_{bc}^a$ is the structure constant. 

In order to connect his gauge field $b_{\mu}^a$ to the metric, Yang introduced a higher order curvature term in the action. The final form of his gravitational action is
\begin{eqnarray}
&&S_g(b_\mu^{a},g^{\mu\nu}) \nonumber \\
&&\qquad= \int d^4x \sqrt{-g} \left(g^{\mu\alpha} g^{\nu\beta} C_{ac}^{d} C_{bd}^c f_{\mu\nu}^a f_{\alpha\beta}^b - R^{\alpha\beta\gamma\delta}R_{\alpha\beta\gamma\delta}\right)
,\nonumber\\
&&\label{yang_l}
\end{eqnarray}
Performing variations of the action with respect to the gauge field $b_{\mu}^a$ and the metric, one arrives at two types of equation of motion.
The index $a$ in $b_{\mu}^a$ has $4\times 4$ values and it can be redefined as $a\equiv\{kl\}$ where $k$ and $l$ run from $0$ to $3$. Yang used an ansatz $b_\alpha^{\{kl\}}\delta_{k\mu}\delta_{l\nu} =\{^\mu_{\nu\alpha}\}$ that satisfies the equation derived from the variation of the action with respect to the metric. 
It gives the relation;
\beq
f_{\mu\nu}^{(\alpha\beta)}=-R^\alpha_{~\beta\mu\nu},
\eeq
and the other equations become
\beq
\nabla_{\beta}R_{\mu\alpha}=\nabla_{\alpha}R_{\mu\beta}.
\label{yang_ge}
\eeq
This is Yang's gravitational equation based on the gauge theory and it  covers the solutions of  Einstein equation for pure space, that is, without matter \cite{PhysRevLett.33.445}.

In Stephenson's theory, there are two EOMs which stem from the variations of the action with respect to the metric and the connection, respectively \cite{Steph1958}:
\begin{eqnarray}\nonumber
&-&R^{\mu\alpha\beta\gamma}R_{\nu\alpha\beta\gamma}+R^{\alpha\mu\beta\gamma}
R_{\alpha\nu\beta\gamma}\\
&+&2R^{\alpha\beta\mu\gamma}R_{\alpha\beta\nu\gamma}-\frac{1}{2}g^{\mu}_{\nu}R^{\alpha\beta\gamma\delta}R_{\alpha\beta\gamma\delta}=0,
\label{Stephenson1} 
\end{eqnarray}
\beq
\nabla_{\alpha}(R_{\mu}^{\nu\sigma\alpha}\sqrt{-g})=0.
\label{Stephenson2} 
\eeq 
The relation between the metric and the connection, however, is different from that in GR. In particular, 
the affine connection in his theory 
%is not the same as 
can in principle be different from the Levi-Civita connection. 
%This is in contrast with Yang's theory. 
If one identifies the affine connection as the Levi-Civita connection in Stephenson's theory, the resulting EOMs are not equivalent to Yang's. 
%{\color{red}On the other hand,} if the connection is recognized as the Christoffel symbols, {\color{red}then} 
Then, Stephenson's equations become
\begin{eqnarray}
&&R^{\mu\alpha\beta\gamma}R_{\nu\alpha\beta\gamma}-\frac{1}{4}g^{\mu}_{\nu}R^{\alpha\beta\gamma\delta}R_{\alpha\beta\gamma\delta}=0,
\label{S1}
\\
&&\nabla_{\alpha}R_{\mu}^{\nu\sigma\alpha}=0.
\label{S2}
\end{eqnarray}
These equations can be recognized as Yang's gravitational equation (Eq.(\ref{S1})) under the additional constraint of Eq.(\ref{S2}). In this sense, the solution space of the resultant Stephenson's equations must be a subset of Yang's gravitational equation. Without specifying the connection as Levi-Civita connection, Stephenson's equations should in principle have different solutions from Yang's.

%{\color{blue}
%Cook obtained Eq.(\ref{CamEq}) from his action using the Palatini formalism as Stephenson did. Hence, it is expected that there should be two independent EOMs resulting from the variations with respect to the metric and the connection, respectively. However, Cook obtains only one EOM in his derivation \cite{Cook:2009mx}. In his theory, Eq.(\ref{CamEq}) comes from the variation with respect to the connection. He did not consider the other EOM related to the variation with respect to the metric.  %This is the second problem in Cook's theory.
%}

\section{Problems with Cook's Theory}
\label{apdx:cook}
  %Cook's  theory is to treat gravitation as a gauge theory by suggesting an analogy between the gravitational theory and the Maxwell theory. Based on that, he constructs his theory at the action level where, in addition to Yang's pure gravity term, there is an interaction term that couples the matter current density to the gravitational gauge field, identified as the affine connection. 

Cook introduced the matter action to Stephenson's theory. 
His recipe, however, not only retains the original problem of Stephenson's theory but also introduces another one.
The action he proposed is
\begin{eqnarray}
S_{G}&=&\frac{-1}{16\pi}\int_{\Sigma}(R^{\alpha\beta\mu\nu}R_{\alpha\beta\mu\nu} \nonumber \\
& &\mbox{}+16\pi J_{\mu}^{\ \alpha\beta}\Gamma^{\mu}_{\ \alpha\beta})\sqrt{-g}\ d^{4}x,
\label{Cook:GravAction2}
\end{eqnarray}
where the tensor $J_{\mu}^{\ \alpha\beta}$ is Cook's current density, which has the same form as Camenzind's. This matter action term, however, is not general covariant because the connection is not a covariant tensor \cite{Alder}. 

%The EOMs in this theory will therefore be dependent on inertial frames. 

The lack of general covariance must be closely related to the non-conservation of the current density $J$.
According to Noether's theorem, symmetry property of the action always goes hand-in-hand with the conservation of the current.
Therefore, it seems impossible to construct a general covariant action based on Camenzind's current density.
This suggests that one should search for a different form of the current density other than that of Camenzind's, under the constraint that GR must be recovered. We will investigate this further.

\end{document}